\documentclass[aps,prl,groupedaddress,showpacs,twocolumn]{revtex4-1}
\usepackage[utf8]{inputenc}
\usepackage{hyperref}
\usepackage{graphicx}  % needed for figures
\usepackage{dcolumn}   % needed for some tables
\usepackage{bm}        % for math
\usepackage{amssymb,amsmath}   % for math
\usepackage{uri}

\usepackage[x11names, rgb, html]{xcolor}
\definecolor{sbase03}{HTML}{002B36}
\definecolor{sbase02}{HTML}{073642}
\definecolor{sbase01}{HTML}{586E75}
\definecolor{sbase00}{HTML}{657B83}
\definecolor{sbase0}{HTML}{839496}
\definecolor{sbase1}{HTML}{93A1A1}
\definecolor{sbase2}{HTML}{EEE8D5}
\definecolor{sbase3}{HTML}{FDF6E3}
\definecolor{syellow}{HTML}{B58900}
\definecolor{sorange}{HTML}{CB4B16}
\definecolor{sred}{HTML}{DC322F}
\definecolor{smagenta}{HTML}{D33682}
\definecolor{sviolet}{HTML}{6C71C4}
\definecolor{sblue}{HTML}{268BD2}
\definecolor{scyan}{HTML}{2AA198}
\definecolor{sgreen}{HTML}{859900}
\hypersetup{
  colorlinks = true,
  allcolors = {blue}
}

\newcommand{\Stot      }{S_{\rm tot}}
\newcommand{\DStot      }{\text{d}S_{\rm tot}}
\newcommand{\Dt      }{\text{d}t}

\newcommand{\itos     }{It\^o's }
\newcommand{\kb     }{k_{\rm B}}

\newcommand{\bsigma     }{\mbox{\boldmath$\sigma$}}
\newcommand{\bmu     }{\mbox{\boldmath$\mu$}}

\newcommand{\bD     }{\mbox{\boldmath$D$}}

\usepackage{color}

\begin{document}
\title{Generic Properties of Stochastic Entropy Production}
\author{Simone Pigolotti$^{1,2}$}\email{simone.pigolotti@oist.jp}
\author{Izaak Neri$^{1,3}$}\email{izaakneri@posteo.net}
\author{\'{E}dgar Rold\'{a}n$^{1}$}\email{edgar@pks.mpg.de}
\author{Frank J\"{u}licher$^{1,4}$}\email{julicher@pks.mpg.de}
\affiliation{$^1$ Max Planck Institute for the Physics of Complex Systems,
  N{\"o}thnitzerstra{\ss}e 38, 01187 Dresden, Germany \\
$^2$ Biological Complexity Unit, Okinawa Institute for Science and
  Technology and Graduate University, Onna, Okinawa 904-0495, Japan\\
${^3}$Max Planck Institute of Molecular Cell Biology and Genetics, Pfotenhauerstra{\ss}e 108, 01307 Dresden, Germany\\
${^4}$Center for Systems Biology Dresden, Pfotenhauerstra{\ss}e 108, 01307 Dresden, Germany}

\begin{abstract} 
  We derive an It\^{o} stochastic differential equation for entropy production in
  nonequilibrium Langevin processes.  Introducing a random-time
  transformation, entropy production obeys a one-dimensional drift-diffusion
  equation, independent of the underlying physical model.  This transformation
  allows us to identify generic properties of entropy production. It also
  leads to an exact uncertainty equality relating the Fano factor of entropy production
  and the Fano factor of the random time, which we also generalize
  to non steady-state conditions. \end{abstract} 
\maketitle

The laws of thermodynamics can be extended to mesoscopic
systems~\cite{sekimoto2010stochastic,bustamante2005nonequilibrium,maes2000on,jarzynski2008nonequilibrium,seifert2012stochastic}.
For such systems, energy changes on the order of the thermal energy $k_{\rm
  B}T$ are relevant. Here, $\kb$ is the Boltzmann constant and $T$ the
temperature. Therefore, thermodynamic observables associated with mesoscopic
degrees of freedom are stochastic.  A key example of such thermodynamics
observables is the stochastic entropy production in nonequilibrium
processes. Recent experimental advances in micromanipulation techniques permit
the measurement of stochastic entropy production in the
laboratory~\cite{martinez2017colloidal,gomez2011fluctuations,speck2007distribution,gavrilov2017direct,koski2015chip}.

Certain statistical properties of stochastic entropy production are
generic, i.e., they are independent of the physical details of a
system.  Examples of such generic properties are the celebrated
fluctuation theorems, for reviews
see~\cite{bustamante2005nonequilibrium,jarzynski2008nonequilibrium,seifert2012stochastic}.
Recently, it was shown that 
infima and passage probabilities of entropy production are also generic
\cite{neri2016infimum}. Other statistical properties of entropy
production are system-dependent, such as the mean value~\cite{bo2016multiple,harada2005equality,kawai2007dissipation,parrondo2009entropy}, the
variance~\cite{barato2015thermodynamic,pietzonka2016universal}, 
the first-passage times of entropy
production~\cite{roldan2015decision,saito2016waiting,gingrich2017fundamental} and the
large deviation function~\cite{lebowitz1999gallavotti,mehl2008large}.
Nevertheless, these properties are sometimes constrained by universal
bounds~\cite{neri2016infimum,kawai2007dissipation,barato2015thermodynamic,gingrich2016dissipation,pietzonka2016universal,polettini2016tightening,garrahan2017simple,pietzonka2017finite,gingrich2017proof}.
It remains unclear which statistical properties of stochastic
entropy production are generic, and why.

%Third Paragraph: Statement of the aim of the paper 

In this Letter, we introduce a theoretical framework which addresses
this question for nonequilibrium Langevin processes.  We identify
generic properties of entropy production by their 
independence of a stochastic variable $\tau$ which we call {\em
  entropic time}.  We find that the evolution of steady-state entropy
production as a function of $\tau$ is governed by a simple
one-dimensional drift-diffusion process, independent of the
underlying model. This allows us to identify a set of generic
properties of entropy production and obtain exact results
characterizing entropy production fluctuations.

\begin{figure}[htb]
\centering
\includegraphics[width=.45\textwidth]{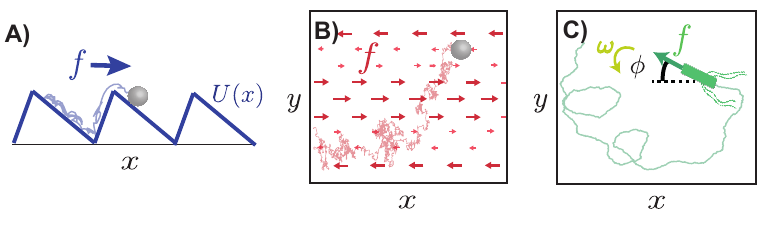}
\caption{Examples of nonequilibrium steady states. (A) Brownian particle
  driven by a constant non-conservative force in a periodic 1D sawtooth potential,
 $\text{d}X/\text{d}t=\mu[f-\partial_X U(X)]+\sqrt{2D}\xi$, with the
 potential $U(x)=(U_0\,x)/x^*$ for $x\in[0,x^*]$ and $U(x)=U_0 (1-x)/(1-x^*)$ for
 $x\in[x^*,1]$. (B)~2D transport in a force field: $\text{d}X/\text{d}t=\mu
 f\cos(2\pi Y)+\sqrt{2D}\xi_x$ and $\text{d}Y/\text{d}t=\sqrt{2D}\xi_y$. (C)
 Chiral active Brownian motion described by 3 degrees of freedom: position coordinates
 $\text{d}X/\text{d}t=\mu f \cos(\phi)+\sqrt{2D}\xi_x$,
 $\text{d}Y/\text{d}t=\mu f \sin(\phi)+\sqrt{2D}\xi_y$ and orientation angle
 $\text{d}\phi/\text{d}t=\mu_\phi
 \omega+\sqrt{2D_\omega}\xi_\omega$. In (B) and (C) $U=0$ and $f$ is
 an external non-conservative force.
 \label{fig:models}}
\end{figure}

We consider a mesoscopic system described by $n$ slow degrees of freedom $\vec{X} =
(X_1(t), X_2(t), \ldots, X_n(t))^{\mathsf{T}}$.  The system is in contact with a thermostat at
temperature $T$.  The stochastic dynamics of the system can be described by
the probability distribution $P(\vec{X}, t)$ to find the system in a
configuration $\vec{X}$ at time $t$ . This probability distribution satisfies the
Smoluchowski equation
\begin{equation}
\partial_t P = -\vec\nabla \cdot \vec{J}\quad,
\label{eq:smo}
\end{equation}
where the probability current is given by
\begin{equation}
\vec{J} = \bmu \cdot\vec{F}\:P -\bD\cdot \vec{\nabla} P \quad.
\label{eq:j}
\end{equation}
Here we have introduced the force at time $t$, $\vec{F}=-\vec{\nabla}
U(\vec{X}(t), t) + \vec{f}(\vec{X}(t), t)$, where $U$ is a potential and
$\vec{f}$ is a non-conservative force. We always imply no flux or periodic boundary conditions.  The state-dependent mobility and
diffusion tensors, $\bmu(\vec{X}(t))$ and $ \bD(\vec{X}(t))$ respectively, are
symmetric and obey the Einstein relation $ \bD=k_{\rm B}T \bmu $. This system
can also be represented by a Langevin equation with multiplicative noise 
as~\cite{lebowitz1999gallavotti,maes2008steady}
\begin{eqnarray}
\label{eq:langevin}
\frac{{\rm d}\vec{X}}{{\rm d}t} = \bmu \cdot \vec{F}+ 
\vec{\nabla}\cdot \bD + \sqrt{2}\:\bsigma\cdot\vec{\xi}\quad.  \label{eq:Ito}
\end{eqnarray}
Here $\vec{\xi}(t)=(\xi_1(t),\xi_2(t),\dots,\xi_n(t))^{\mathsf{T}}$ is a
Gaussian white noise with mean $\langle {\xi}_i(t) \rangle = {0}$ and autocorrelation
$\langle {\xi_i(t) \xi_j(t') } \rangle=\delta_{ij}\delta(t-t')$ where
$\langle\dots\rangle$ denotes an ensemble average.
Here and throughout the paper the noise terms are interpreted in the It\^{o} sense.
 The tensor $\bsigma$ obeys
$\bsigma \bsigma^{\mathsf{T}}=\bD$ and can be chosen as $\bsigma=\bD^{1/2}$.
In the  It\^{o} interpretation, the term $\vec{\nabla}\cdot \bD$ is required for
consistency with Eqs.~(\ref{eq:smo}) and \eqref{eq:j} as it compensates a
noise-induced drift~\cite{lau2007state}. Examples of systems described by
Eq.~\eqref{eq:langevin} that we consider in this paper are represented in
Fig.~\ref{fig:models}: a colloidal particle driven by a constant force along a
one-dimensional periodic potential (Fig.~\ref{fig:models}A); a colloidal
particle in a two-dimensional non-conservative force field pointing in the $x$
direction (Fig.~\ref{fig:models}B); and a chiral active Brownian motion in two
dimensions \cite{bechinger2016active} (Fig.~\ref{fig:models}C).

We now discuss the stochastic thermodynamics of the process described by
Eq.~\eqref{eq:Ito}. In \itos calculus, the rate of change of the potential
$U(\vec{X}(t),t)$ is given by \itos lemma~\cite{oksendal2013stochastic}:
  \begin{equation}
\frac{\text{d}U}{\text{d}t} = \partial_t U +\vec{\nabla} U(\vec{X}(t), t)  \cdot \frac{\text{d}\vec{X}}{\text{d}t}  + {\rm{Tr}}\left[\bD\cdot \vec{\nabla}\, \vec{\nabla} U\right]
\label{eq:Uito}
 \end{equation}
 where Tr denotes the trace and the dots denote  tensor contractions.  In stochastic thermodynamics, the
 first law can be expressed as
 $\text{d}U=\text{d}W+\text{d}Q$, where $\text{d}W$ is the work performed on
 the system and $\text{d}Q$ is the mesoscopic heat exchanged with the
 thermostat during a time interval $\text{d}t$~\cite{sekimoto2010stochastic}. In \itos calculus, the rates of change of work
 and heat are given by~\cite{lebowitz1999gallavotti}
 \begin{eqnarray}
 \frac{{\rm d}W}{\text{d}t}  &=& \partial_t U+ \vec{f}\cdot  
\frac{{\rm d}\vec{X}}{\text{d}t} + {\rm{Tr}}\left[\bD\cdot 
\vec{\nabla}\vec{f}\right]
 \label{eq:Wito}\\
\frac{{\rm d}Q}{\text{d}t} &=& -  
\vec{F} \cdot \frac{{\rm d}\vec{X}}{\text{d}t} -   
{\rm{Tr}}\left[\bD  \cdot\vec{\nabla}\,\vec{F}\right] \quad.
\label{eq:Qito}
 \end{eqnarray}
 The expressions~\eqref{eq:Wito} and~\eqref{eq:Qito} are the It\^{o} versions of the stochastic work and mesoscopic heat originally
 defined by Sekimoto using the Stratonovich
 interpretation~\cite{sekimoto1998langevin,sekimoto2010stochastic}.

 We define the stochastic entropy production $S_{\rm tot}/\kb$ as the
 logarithm of the ratio of probabilities of forward and time-reversed
 stochastic trajectories~\cite{lebowitz1999gallavotti,maes2000on,seifert2005entropy}. This definition is
 equivalent to $\DStot/\Dt= -(1/T)\text{d}Q/\text{d}t -\kb \text{d}\ln
 P(\vec{X}(t),t)/\text{d}t$, where the first term can be interpreted as an exchange
 of entropy with the reservoir and the second term as a change of system
 entropy. Using Eq.~\eqref{eq:Qito} and \itos lemma, as in Eq.~\eqref{eq:Uito}
 (see \cite{footnote_suppl}), we obtain the following It\^{o} stochastic differential
 equation for the entropy production rate
    \begin{eqnarray}
      \frac{{\rm d}S_{\rm tot}}{{\rm d}t} =  
      -2k_{\rm B}\partial_t \ln P +  v_{\rm S} + 
\sqrt{2k_{\rm B}v_{\rm S}}\:\xi_S\quad.  \label{eq:entropy}
 \end{eqnarray}
 \begin{figure}[ht]
\centering
\includegraphics[width=6cm]{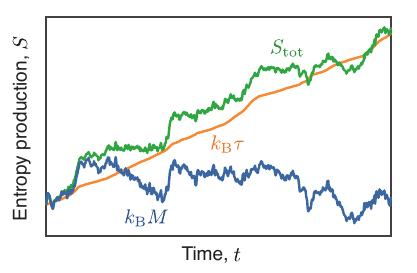}
\caption{Illustration of the decomposition of stochastic entropy production. In nonequilibrium steady states, the stochastic entropy production $S_{\rm tot}(t)/\kb$ (green) is given by the sum of the monotonously increasing entropic time $\tau(t)$ (orange), and the martingale process $M(t)$ (blue), see Eq.~\eqref{eq:doob}.
%Illustration of the decomposition of stochastic entropy production given by Eq.~\eqref{eq:doob}. In nonequilibrium steady states, the stochastic entropy production $S_{\rm tot}(t)$ (green) is given by the Boltzmann constant $k_{\rm B}$ times the sum of the monotonously increasing entropic time $\tau(t)$ (orange), see Eq.~\eqref{eq:tau}, and the martingale process~$M(t)= \sqrt{2/\kb} \int_0^t \sqrt{v_S(\vec{X}(t'))} \xi_S(t') \text{d}t'$ (blue),  see Eq.~\eqref{eq:entropy}, with $v_S$ given by Eq.~\eqref{eq:vs}.
 \label{fig:sketch}}
\end{figure}
Here we define the entropic  drift  $v_S(\vec{X}(t),t)\geq 0$ as 
 \begin{eqnarray}
     v_S =  k_{\rm B}\frac{\vec{J}\cdot \bD^{-1}\cdot\vec{J}}{P^2}\quad,
     \label{eq:vs}
 \end{eqnarray}
 which on average equals the average rate of entropy production, $ \langle
 v_S\rangle=\langle \text{d}S_{\rm tot}/\text{d}t\rangle
 $~\cite{maes2008steady,seifert2012stochastic}. 
 Entropy fluctuations are governed by the noise term~$\xi_S = \vec{\xi}\cdot
\bsigma^{-1}\cdot \vec{J}/\sqrt{\vec{J} \cdot \bD^{-1}\cdot \vec{J}}$ which is a
one-dimensional Gaussian white noise with $\langle\xi_S (t)\rangle=0$ and
$\langle \xi_S(t)\xi_S(t')\rangle=\delta(t-t')$.  The It\^{o}
Eq.~\eqref{eq:entropy} is equivalent to the Langevin equation for entropy
production in the Stratonovich interpretation given in
Ref.~\cite{seifert2005entropy}.  For each trajectory generated by
Eq.~\eqref{eq:Ito}, Eq.~\eqref{eq:entropy} generates the corresponding entropy
production.  From Eq.~\eqref{eq:entropy} we can derive several generic
properties of stochastic entropy production in nonequilibrium processes.

We first discuss
 properties of nonequilibrium steady states for which $\partial_t P
 =0$. We now calculate the time derivative of $e^{-S_{\rm
     tot}/k_{\rm B}}$ in steady state. Using \itos lemma, we obtain from Eq.~\eqref{eq:entropy}
\begin{equation}
\frac{\text{d}e^{-S_{\rm tot}/k_{\rm B}}}{\text{d}t}=
-\sqrt{2 \frac{v_S}{k_{\rm B}}}\, e^{-S_{\rm tot}/k_{\rm B}}\,\xi_S\quad, 
\label{eq:gbm}
\end{equation}
which reveals that $e^{-S_{\rm tot}/k_{\rm B}}$ is a geometric Brownian motion
with zero drift and time-dependent diffusion coefficient.  The fact that
$e^{-S_{\rm tot}/k_{\rm B}}$ has no drift implies that $e^{-S_{\rm tot}/k_{\rm
    B}}$ is a martingale
process~\cite{oksendal2013stochastic,chetrite2011two,neri2016infimum}.  Using
$S_{\rm tot}(0)=0$ the integral fluctuation theorem $\langle e^{-S_{\rm
    tot}(t)/k_{\rm B}}\rangle =1$ follows immediately from Eq.~\eqref{eq:gbm}.

\begin{figure*}[htb]
\includegraphics[width=\textwidth]{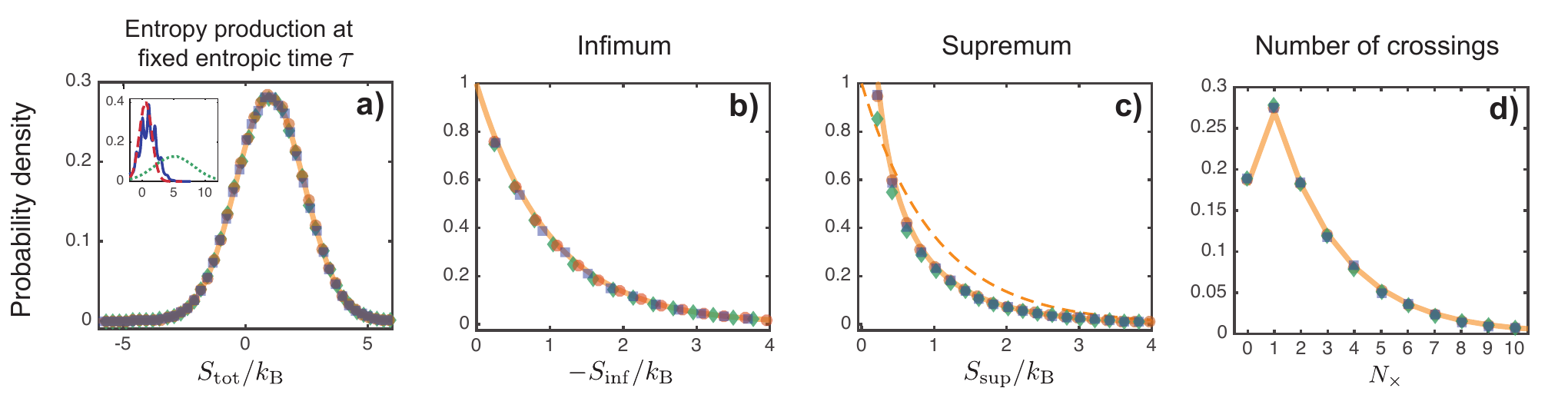}
\caption{Generic properties of stochastic entropy production.  Distributions
  of a) entropy production at fixed $\tau=1$, (b) infimum of entropy
  production, c) supremum of entropy production before the infimum, d) number
  of crossings of entropy production, with $\Delta=0.2\kb$. The symbols are obtained from numerical
  simulations of the three models sketched in Fig. 1 (blue squares, model A;
  red circles, model B; green diamonds, model C). The inset of a) shows
  numerically estimated distributions of $S_{\rm tot}$ for the three models at
  fixed $t=1$ for comparison. The solid orange curves are the theoretical
  expressions a) a Gaussian distribution with average $\kb\tau$ and variance
  $2\kb^2\tau$ b) an exponential distribution with average $-\kb$ c)
  Eq.~\eqref{eq:supremum}; d) Eq.~\eqref{eq:crossings}. The dashed line in c)
  is the theoretical distribution of minus the infimum for comparison. In all
  simulations, parameters are $\bmu=\mathbb{I}$, $\bD=\kb T~\mathbb{I}$, where
  $\mathbb{I}$ is the identity matrix, and $f=1$. In Model A we chose
  $U_0=\kb T$ and $x^*=0.3$. In model C we chose $\omega=2$. Here and in the following
  figures, each point represents an average over $10^6$
  simulations.  \label{fig:crossings}}
\end{figure*}

In steady state, Eq.~\eqref{eq:entropy} can be simplified by introducing the
dimensionless {\em entropic time}%~
\begin{equation}\label{eq:tau}
  \tau=\frac{1}{k_{\rm B}}\int_0^t v_S(\vec{X}(t'))\,\text{d}t' \quad,
\end{equation}
which is an example of a random time~\cite{oksendal2013stochastic}. Note
that, in steady state, $v_S(\vec{X}(t),t)=v_S(\vec{X}(t))$ represents the expected rate of entropy production at a
given point in phase space $\vec{X}(t)$ and $\tau$ thus represents the accumulated expected entropy production.    In nonequilibrium situations with $v_S>0$,
the entropic time $\tau(t)$ is monotonously increasing with $t$. Integrating
Eq.~\eqref{eq:entropy} we obtain
\begin{equation}\label{eq:doob}
S_{\rm tot}(t)/\kb=\tau(t) +  M(t)\quad .
\end{equation}
Equation~\eqref{eq:doob} represents the decomposition of entropy production into a
monotonously increasing process $\tau(t)$ and a martingale $M(t)=\sqrt{2/\kb}\int_0^t \sqrt{v_S(\vec{X}(t'))} \xi_S(t') \text{d}t'$ that has zero mean, $\langle
M(t)\rangle =0$, as illustrated in Fig.~\ref{fig:sketch}. This decomposition
is unique and is known as the Doob-Meyer
decomposition~\cite{liptser2013statistics}.

We now discuss an important implication of Eqs.~\eqref{eq:entropy}
and~\eqref{eq:tau}. 
Performing the random-time transformation $t\to\tau$ in Eq.~\eqref{eq:entropy}
we obtain a Langevin equation for steady state entropy production at entropic times \cite{oksendal2013stochastic}
\begin{eqnarray}\label{eq:Stau}
\frac{1}{k_{\rm B}}\frac{{\rm d}S_{\rm tot}}{{\rm d}\tau} =   1 +    \sqrt{2}\:\eta(\tau)\quad,
\end{eqnarray} 
where $\eta(\tau(t)) = \sqrt{k_{\rm B}\,/\,v_S(\vec{X}(t))}\,\xi_S(t)$ such that
$\eta(\tau)$ is Gaussian white noise with $\langle\eta(\tau)\rangle=0$ and
$\langle\eta(\tau)\eta(\tau')\rangle=\delta(\tau-\tau')$.
Equation~\eqref{eq:Stau} states that a temporal trajectory of entropy production of
any nonequilibrium steady state can be mapped to a trajectory of a drift-diffusion
process with constant drift $\kb$ and diffusion coefficient $\kb ^2$, where
the mapping consists in a time-dependent, stochastic contraction or dilation of time. This
implies that all properties of $S_{\rm tot}$ that are invariant under such transformation are generic. 

One such property is the distribution of entropy production at fixed values of
$\tau$, which must be a Gaussian with average $\kb\tau$ and variance
$2\kb^2\tau$ because of Eq.~\eqref{eq:Stau}. This is indeed the case for all
three model examples, see Fig.  \ref{fig:crossings}a. Note that the
distribution of entropy production at fixed time $t$ are very different for
the three models, as shown in the inset of Fig. \ref{fig:crossings}a. Another
generic property is the distribution of the global infimum of entropy
production $S_{\rm inf}$, previously derived using martingale
theory~\cite{neri2016infimum} and given by an exponential distribution
$P(S_{\rm inf})=e^{S_{\rm inf}/\kb}/\kb$ with mean $-\kb$ and $S_{\rm
  inf}\le 0$
(Fig.~\ref{fig:crossings}b). Also the supremum of entropy production before
the infimum is generic and distributed according to
\begin{equation}\label{eq:supremum}
P(S_{\rm sup})=2 e^{S_{\rm sup}/k_{\rm B}}
\mathrm{acoth}(2 e^{S_{\rm sup}/k_{\rm B}}-1)-1\quad
\end{equation}
with $S_{\rm sup}\ge 0$. Its average value is $\langle S_{\rm sup}\rangle=(\pi^2/6-1)k_{\rm B}\approx
0.645 k_{\rm B}$.  The number of times that entropy production crosses a given threshold value is also generic.  An example is the number of times $N_{\!\times}$ that entropy production crosses from
$-\Delta$ to $\Delta$ with $\Delta>0$. The
distribution of $N_{\!\times}$ is
\begin{equation}\label{eq:crossings}
P(N_{\!\times};\Delta)=
\left\{
\begin{array}{lr}
1-e^{-\Delta/k_{\rm B}} & N_{\!\times}= 0 \\
2\sinh ( \Delta/k_{\rm B}) e^{-2 N_{\!\times}\Delta/k_{\rm B}} & N_{\!\times}\ge 1 
\end{array}
\right.
\end{equation}
Equation~\eqref{eq:supremum} and \eqref{eq:crossings} are in excellent agreement with
numerical simulations, as shown in Fig.~\ref{fig:crossings}c,d.

With equation~\eqref{eq:doob}, we can also compute the moments of $S_{\rm
  tot}(t)$. The first moment reads simply $\langle S_{\rm tot}\rangle =\kb\langle
\tau\rangle$. The second moment is $\langle S^2_{\rm tot} \rangle = 2 \kb^2
\langle \tau \rangle+ \kb^2\langle\tau^2\rangle$, see
\cite{footnote_suppl}. Combining these two results, the Fano factor of the
entropy production can be expressed as
\begin{equation}\label{eq:uncertainty}
\frac{1}{k_{\rm B}}\frac{\sigma^2_{S_{\rm tot}}}{\langle S_{\rm tot}\rangle}  
= 2 + \frac{\sigma^2_{\tau}}{\langle \tau\rangle}\quad,
\end{equation}
where $\sigma^2_y=\langle y^2\rangle-\langle y\rangle^2$ denotes the variance.
\begin{figure}[h]
\includegraphics[width=6.5cm]{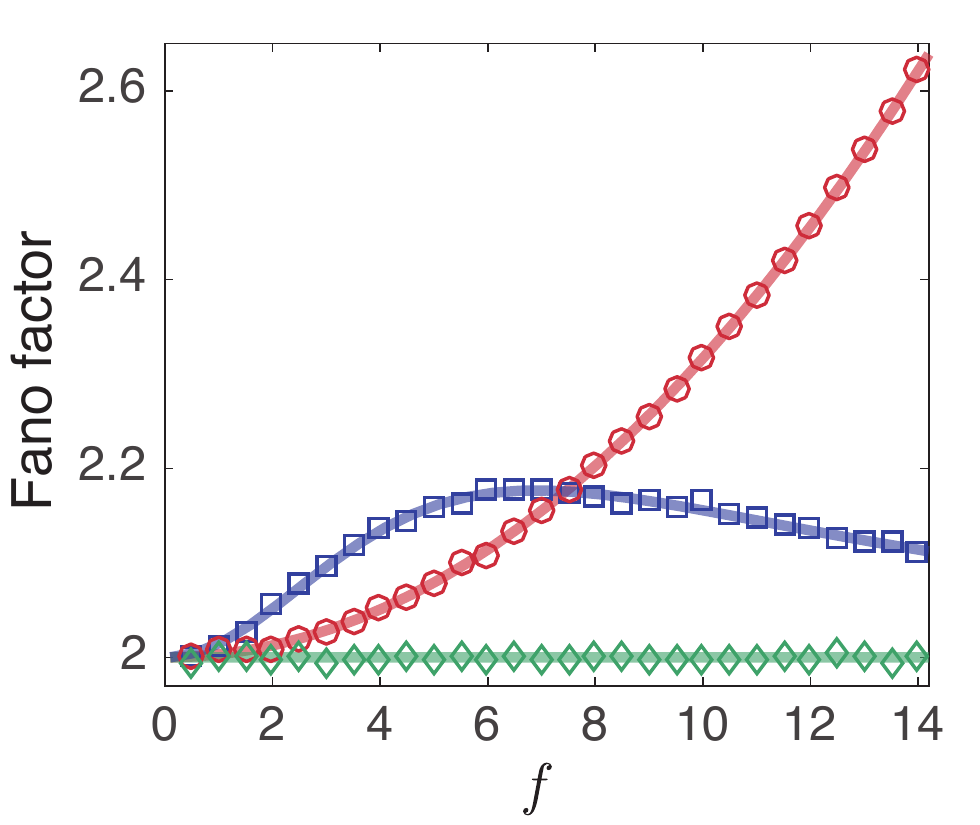}
\caption{Thermodynamic Fano factor equality. Long-time Fano factor of entropy
  production $ \sigma^2_{\Stot}/(\kb \langle \Stot\rangle)$ as a function of
  the external force $f$. The symbols are obtained from numerical
  simulations of the models shown in Fig.~\ref{fig:models}A (blue),
  Fig.~\ref{fig:models}B (red), and Fig.~\ref{fig:models}C (green). The solid
  lines are the prediction of Eq.~\eqref{eq:uncertainty} and have been
  calculated by means of Eq.~\eqref{greenkubo} (see
  \cite{footnote_suppl}). All the parameters of the
  numerical simulations except of the external force $f$ are the same as in
  Fig.~\ref{fig:crossings}. \label{fig:fano}}
\end{figure}
The thermodynamic {\em Fano factor equality} given by Eq.~\eqref{eq:uncertainty}
is an exact relation, valid for finite times, between the fluctuations of
entropy production and the fluctuations of the entropic time $\tau$.  This
equation provides further physical insight into the previously introduced
finite-time uncertainty relation,~$\sigma^2_{S_{\rm tot}}/\langle S_{\rm
  tot}\rangle\geq 2\kb$~\cite{pietzonka2017finite,gingrich2017proof}. The variance obeys the equality $\sigma^2_{S_{\rm
    tot}}/\langle S_{\rm tot}\rangle= 2\kb$ only if the entropic
time satisfies $\sigma_\tau^2/\langle\tau\rangle=0$, which holds e.g. near equilibrium. In this case, the distribution of entropy
production is Gaussian. Another example for which $\sigma_\tau^2/\langle\tau\rangle=0$ is the chiral active Brownian motion shown in Fig.~\ref{fig:models}c.

For long times, the variance of the entropic time can be estimated by a
Green-Kubo formula as an integral over a correlation function  \cite{footnote_suppl}
\begin{equation}\label{greenkubo}
\frac{\sigma^2_\tau}{\langle \tau\rangle}
=\frac{2}{\kb\langle v_S\rangle}\int_0^\infty \!\text{d}t'
\left(\langle\, v_S(\vec{X}(t'))\,v_S(\vec{X}(0))\,\rangle - 
\langle v_S \rangle^2\right)\,. 
\end{equation}
 Using Equations~\eqref{eq:uncertainty} and~\eqref{greenkubo} we
 obtain explicit expressions for the  Fano factor as a function of the
 driving force for our three models, see Fig.~\ref{fig:fano} for a
 comparison with numerical simulations.

\begin{figure}[h]
\includegraphics[width=5.5cm]{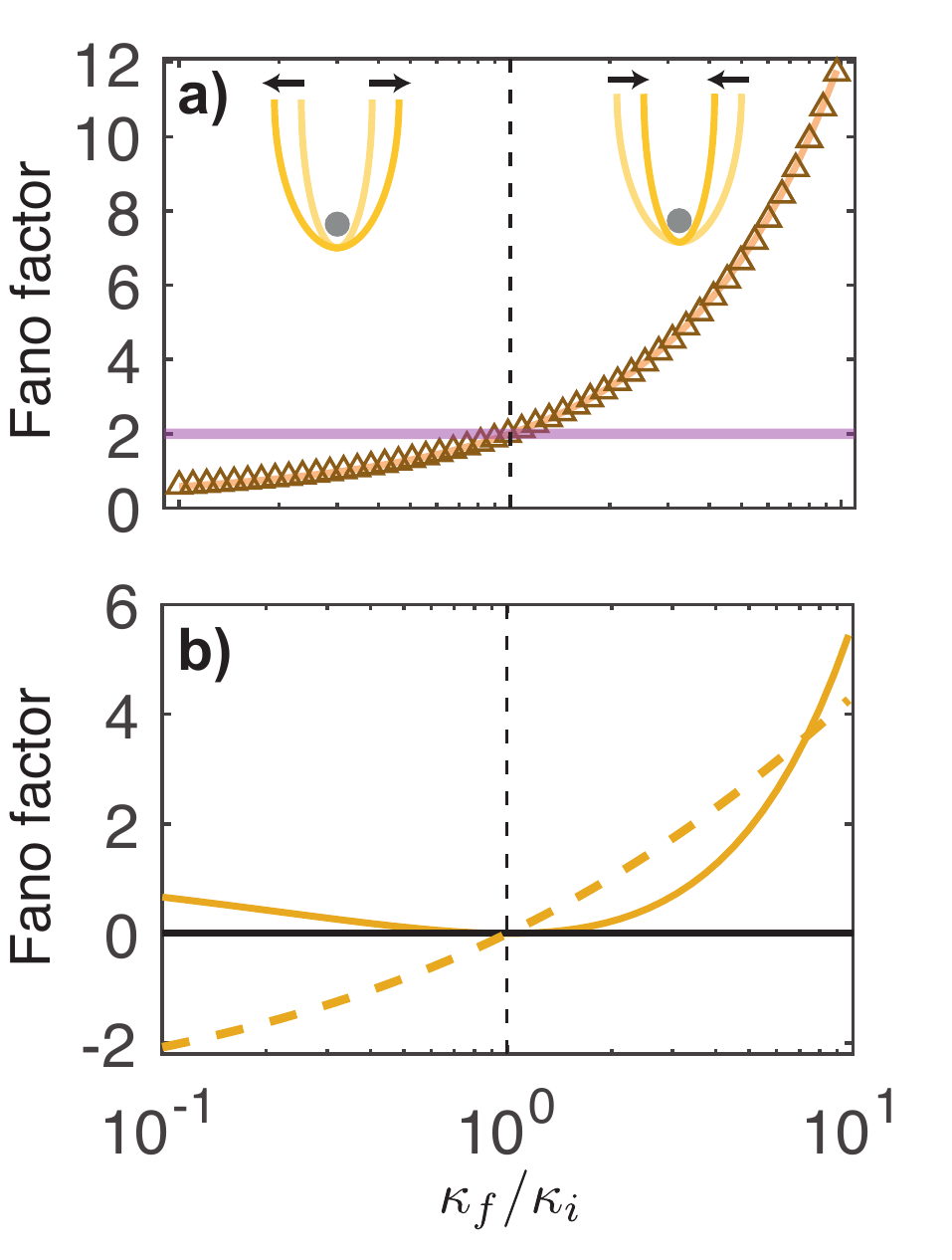}
\caption{Fano factor of stochastic entropy production out of steady state.
The position of a Brownian particle is governed by the equation
  ${\rm d}X/{\rm d}t=-\mu\kappa_f X+\sqrt{2D}\xi$ with $\mu=1$ and
  $D=\mu \kb T$. The particle is initially at equilibrium with stiffness
  $\kappa_i=1$ (see inset).  a) Comparison
  between the exact value (orange line) of the long-time Fano factor of entropy production and the value obtained from numerical simulations (brown triangles) as a function of $\kappa_f$. 
  The exact value is given by $\sigma^2_{S_{\rm tot}}/\kb\langle S_{\rm tot}\rangle=(\kappa_f/\kappa_i-1)^2/[(\kappa_f/\kappa_i-1)-\log(\kappa_f/\kappa_i)]$ see Supplemental Material for details~\cite{footnote_suppl}. 
 In simulations we measure $\langle \tau\rangle$, $\sigma^2_{\tau}$ and $\Omega$ and use Eq.~\eqref{eq:uncertainty2}.
    The horizontal purple line is set to $2$ for comparison. b) Behaviour of $\sigma^2_{\tau}/\langle \tau\rangle$ (solid line) and $2\Omega/\langle\tau\rangle$ (dashed line) as a  function of $\kappa_f$.   \label{fig:quench}}
\end{figure}

Our theory can also be applied
to nonequilibrium processes out of steady
state.  From Eq.~\eqref{eq:entropy} we derive the general Fano factor equality
\begin{eqnarray}\label{eq:uncertainty2}
\hspace{-0.5cm}  \frac{1}{k_{\rm B}}\frac{\sigma^2_{S_{\rm tot}}}{\langle S_{\rm tot}\rangle}  
 & =& 2 + \frac{\sigma^2_{\tau}}{\langle \tau\rangle}
+\frac{2\Omega}{\langle\tau\rangle} 
\end{eqnarray}
where 
\begin{equation}\label{eq:omega}
\Omega=\frac{1}{\kb}\int_0^t \text{d}t' \int_0^{t'}\!\text{d}t'' \langle\,
-2\partial_{t''}\!\ln P(X(t''),t'')\,v_S(\vec{X}(t'),t')\,\rangle\, ,
\end{equation}
and $\tau=(1/\kb)\int_0^t v_S(\vec{X}(t'),t')\text{d}t'$ is the entropic time for non-steady state processes.
At steady state, $\Omega=0$, and Eq.~\eqref{eq:uncertainty2} reduces to
Eq.~\eqref{eq:uncertainty}.  Note that the argument of the integral in
Eq.~\eqref{eq:omega} is the correlation of the two drift terms in
Eq.~\eqref{eq:entropy} at different times.  In Fig.~\ref{fig:quench}, we
illustrate Eq.~\eqref{eq:uncertainty2} for a particle confined in a harmonic
trap, where the stiffness of the trap is instantaneously quenched from a value
$\kappa_i$ to a value $\kappa_f$. When $\kappa_f>\kappa_i$, one has
$\Omega>0$, so that the Fano factor of entropy production is larger than two
according to Eq.~\eqref{eq:uncertainty2}.  When instead $\kappa_f<\kappa_i$,
one has $\Omega<0$, and the Fano factor of entropy production is lower
than two.

For nonequilibrium processes starting at thermal equilibrium and undergoing a
defined protocol to a final state, one has $TS_{\rm tot}=W-\Delta F$, where
$W$ is the work performed during the protocol and $\Delta F$ is the change of
equilibrium free energy~$F=\langle U\rangle_{\rm eq}+\kb T \langle\, \ln P\,
\rangle_{\rm eq}$ associated with the final and initial
states~\cite{jarzynski1997nonequilibrium,crooks1999entropy,bochkov1977general}.
Here $\langle\,\cdot\, \rangle_{\rm eq}$ denotes an equilibrium average over
the Boltzmann distribution.  For such protocols, Eq.~\eqref{eq:uncertainty2}
implies
%\begin{eqnarray}\label{eq:F}
%\Delta F &=& \langle W\rangle-\frac{\sigma^2_W}{2\kb T}\\
%&+&\frac{\kb T}{2}\sigma^2_\tau  -2\kb T\int_0^t \text{d}t' \text{Cov}[\tau(t');\partial_{t'}\!\ln P(X(t'),t')]\; .\nonumber
%\end{eqnarray}
\begin{eqnarray}\label{eq:F}
\Delta F = \langle W\rangle-\frac{\sigma^2_W}{2\kb T}+\frac{\kb T}{2}\left(\sigma^2_\tau  +2\Omega\right) .
\end{eqnarray}
Note that $\Delta F$ also obeys Jarzynski's equality~$\Delta F=-\kb T \ln
\langle e^{-W/\kb T}\rangle$~\cite{jarzynski1997nonequilibrium}, which has the
form of a cumulant generating function. Comparing it with Eq.~\eqref{eq:F},
one can relate the term in parenthesis in~\eqref{eq:F} to a sum of cumulants of $W/\kb
T$ of order three and higher.  This sum vanishes if the work distribution is
Gaussian~\cite{jarzynski1997nonequilibrium}.

We have shown that, in steady-state Langevin processes,
entropy production is governed by a Langevin equation which only depends on
the system's details via the entropic drift $v_S$. As a consequence all
system-specific features of stochastic entropy production can be absorbed into
a single stochastic quantity, the entropic time $\tau$.  Entropy productions
of different systems at equal entropic time have the same statistics, and all
properties independent of the entropic time are generic.  Fluctuations of the
entropic time uniquely determine the Fano factor of entropy production,
providing physical insight for previously obtained bounds
\cite{barato2015thermodynamic,gingrich2016dissipation,pietzonka2016universal,polettini2016tightening,garrahan2017simple,pietzonka2017finite,gingrich2017proof}.

We have demonstrated our results for coupled overdamped Langevin equations but
expect our results to hold more generally for continuous processes, as is the
case for the infimum of entropy production~\cite{neri2016infimum}. Using the
Doob-Meyer decomposition of entropy production, our definition of entropic
time can also be generalized to underdamped
systems~\cite{celani2012anomalous,ge2014time} and jump processes~\cite{gaspard2004time}.  Our results
can be experimentally tested for example with optical
tweezers~\cite{gomez2011fluctuations,speck2007distribution,martinez2017colloidal,krishnamurthy2016micrometre,argun2016non},
feedback traps~\cite{gavrilov2017direct}, single-electron
transistors~\cite{koski2015chip} and light-activated phototactic microparticles~\cite{lozano2016phototaxis}.

\begin{acknowledgments}
We thank  AC Barato for stimulating  discussions  and A Mazzino and A Vulpiani for suggesting the example of model B.
\end{acknowledgments}

\clearpage

\onecolumngrid

\section{Supplemental Material}
This document provides additional information for the manuscript
``Generic Properties of Stochastic Entropy Production''. It is
organized as follows. Section S1 sketches the
derivation of the It\^{o} stochastic differential equation for the
entropy production. Section S2 presents a derivation of
the Fano-factor equality for steady-state processes. Sections
S3A, S3B, and S3C detail the
calculations of the Fano factor of entropy production for the three
steady-state models discussed in the Main Text. Section
S4 presents the derivation of the Fano-factor equality
out of steady state. Section S5 describes details on the non-equilibrium process shown in Fig.~\ref{fig:quench} of the Main text.

\section{S1.$\quad$Ito stochastic differential equation for entropy production}\label{sec:eqentropy}

In this section, we sketch the derivation of the evolution for the
stochastic entropy production (Eq.~\eqref{eq:entropy} in the Main Text).  We recall that the
rate of total entropy production change can be decomposed into the
rates of system-entropy change and heat change~\cite{seifert2005entropy}
\begin{eqnarray}
\frac{{\rm d}S_{\rm tot}}{{\rm d}t} =  \frac{{\rm d}S_{\rm sys}}{{\rm d}t}  -\frac{1}{T} \frac{{\rm d}Q}{{\rm d}t} \quad,
\end{eqnarray}
with $S_{\rm sys} = -k_B \ln P$ the system entropy.  We express the rate of heat change
as
  \begin{eqnarray}
    -\frac{1}{\kb T}\frac{{\rm d}Q}{{\rm d }t}
      &=&    \frac{\vec{F}}{\kb T} \cdot \frac{{\rm d}\vec{X}}{{\rm d}t} +  \frac{1}{\kb T} {\rm{Tr}}\left[\bD  \cdot\vec{\nabla}\,\vec{F}\right]\nonumber\\ 
    &=& \frac{1}{\kb T}\vec{F} \cdot (\bmu \cdot \vec{F}+\vec{\nabla}\cdot \bD)  + {\rm{Tr}}\left[\bmu  \cdot\vec{\nabla}\,\vec{F}\right] + \frac{1}{\kb T}\sqrt{2}\;\left(\vec{F} \cdot\:\bsigma\cdot\vec{\xi}\right)\quad,\label{eq:Senv}
    \end{eqnarray} where we have used the Einstein relation $\bD=\kb T\bmu$.  We use   It\^o's lemma~\cite{oksendal2013stochastic} to find the following expressions for the rate of system-entropy change 
\begin{eqnarray}
\lefteqn{k^{-1}_{\rm B}\frac{{\rm d}S_{\rm sys}}{ {\rm d}t  }= - \frac{{\rm d}}{{\rm d}t}\ln P}
\nonumber\\
&=&  - \frac{\partial_t P}{P}-   \frac{\vec{\nabla}P}{P}\cdot \frac{{\rm d}\vec{X}}{{\rm d}t}
+  {\rm Tr}\left[\bD\cdot\left[\frac{\vec{\nabla} P \vec{\nabla}P  }{(P)^2} -  \frac{\vec{\nabla}\vec{\nabla} P }{P}\right]  \right]
 \nonumber \\ 
&= & - \frac{\partial_t P}{P} -\frac{\vec{\nabla} P }{P} \cdot  (\bmu\cdot \vec{F} + \vec{\nabla}\cdot \bD) + {\rm Tr}\left[\bD\cdot \frac{\vec{\nabla}P\vec{\nabla} P }{(P)^2}\right]  -{\rm Tr}\left[\bD\cdot \frac{\vec{\nabla}\vec{\nabla} P }{P}\right]  - \sqrt{2}\:{\rm Tr}\left[ \bsigma^T \cdot \frac{\vec{\nabla}P }{P} \vec{\xi} \right]\label{eq:S4} \quad.\nonumber\\
     \end{eqnarray}
The Fokker-Planck equation can be rewritten as 
\begin{eqnarray}
\lefteqn{{\rm Tr}\left[\bD\cdot \frac{\vec{\nabla}\vec{\nabla} P }{P}\right] +(\vec{\nabla}\cdot \bD)\cdot\frac{\vec{\nabla} P }{P}   }&&
\nonumber\\ 
&&= \frac{\partial_t P}{P}  + {\rm Tr}\left[\bmu \cdot \left(\vec{F}\frac{\vec{\nabla} P }{P}\right)\right] +  {\rm{Tr}}\left[\bmu  \cdot\vec{\nabla}\,\vec{F}\right]  + \frac{1}{T}\vec{F}\cdot (\vec{\nabla}\cdot\bD)   \quad. \label{eq:fp}
\end{eqnarray}
After substituting equation~(\ref{eq:fp}) into the expression for the rate of system-entropy change, Eq.~(\ref{eq:S4}), and adding the rate of heat change, Eq.~(\ref{eq:Senv}),  we obtain the following compact expression for the rate of entropy production change
\begin{eqnarray}
\frac{{\rm d}S_{\rm tot}}{{\rm d}t} =  
-2 k_{\rm B} \partial_t \ln P  +  v_{\rm S} +    \sqrt{2k_{\rm B}v_{\rm S}}\:\xi_S\quad,\label{eq:entropySI}
 \end{eqnarray}
 with $v_S = k_{\rm B}\vec{J}\cdot \bD^{-1}\cdot\vec{J}/P^2$, $\xi_S =
 \vec{\xi}\cdot \bsigma^{-1}\cdot \vec{J}/\sqrt{\vec{J} \cdot \bD^{-1}\cdot \vec{J}}$,
 and the probability currents $\vec{J} = \bmu \cdot\vec{F}\:P -\bD\cdot
 \vec{\nabla} P$.

\section{S2.$\quad$Fano Factor of entropy production}\label{sec:fano}

In this section we derive the Fano-factor equality for
stochastic entropy production $S_{\rm tot}$ at finite times
\begin{equation}
\frac{\sigma^2_{S_{\rm tot}}}{\kb\langle S_{\rm tot}\rangle}=
2+\frac{\sigma^2_{\tau}}{\langle \tau\rangle}\quad. \label{eq:uncertaintySI}
\end{equation}
We show that~\eqref{eq:uncertaintySI}  holds for steady state processes $\vec{X}(t)$  satisfying the Langevin
Eq.~\eqref{eq:Ito} in the Main Text.  The Langevin equation for entropy production (\ref{eq:entropySI}) implies that the steady-state stochastic entropy production $S_{\rm
  tot}(t)$ can be expressed as
\begin{equation}
\frac{S_{\rm tot}(t)}{\kb}= \tau(t)+M(t)\quad, \label{eq:s}
\end{equation} 
where $\tau(t)=k^{-1}_{\rm B}\int_0^t v_S(\vec{X}(t'))\text{d}t'$, $M(t)=\sqrt{2/\kb}\int_0^t \text{d}t'\sqrt{v_S(\vec{X}(t'))}\,\xi_S(t')$
and we recall that $v_S(\vec{X}) = k_{\rm B}\vec{J}\cdot \bD^{-1}\cdot\vec{J}/P^2$.   Taking the average of Eq.~\eqref{eq:s},
we find  
\begin{equation}\label{firstmoment}
\frac{\langle S_{\rm tot}(t)\rangle}{\kb} =\langle
\tau(t)\rangle =\frac{t\langle v_S(\vec{X}(t))\rangle}{\kb} \quad.
\end{equation}
From Eq.~(\ref{eq:s}) we find for the second moment of stochastic entropy production
\begin{eqnarray}\label{eq:s2}
  \frac{\langle S^2_{\rm tot}(t)\rangle}{\kb^2} 
&=&\left\langle \left[\frac{1}{\kb}\int_0^t\ \text{d}t'\
  v_S(\vec{X}(t'))+\sqrt{\frac{2}{\kb}}\int_0^t\ \text{d}t' \xi_S(t')\
    \sqrt{v_S(\vec{X}(t'))} \right]^2\right\rangle 
\\
  &=&2\langle \tau(t)\rangle +\langle \tau^2(t) \rangle
  +2\sqrt{\frac{2}{\kb^3}}\left\langle\int_0^t \text{d}t' \int_0^t \text{d}t'' \xi_S(t'')\ v_S(\vec{X}(t'))
    \sqrt{v_S(\vec{X}(t''))} 
  \right\rangle\quad,\nonumber
\end{eqnarray}
where the contribution $2\langle \tau(t)\rangle$ has been obtained using Ito's isometry~\cite{oksendal2013stochastic}. Subtracting $\langle \tau\rangle^2$ from both sides, further dividing by $\langle \tau \rangle$ and using Eq.~\eqref{firstmoment} we obtain
\begin{eqnarray}\label{eq:s3}
\frac{\sigma^2_{S_{\rm tot}}}{\kb\langle S_{\rm tot}\rangle} =
2+\frac{\sigma^2_{\tau}}{\langle \tau\rangle}
  +\frac{2}{\langle\tau\rangle}\sqrt{\frac{2}{\kb^3}}\left\langle\int_0^t \text{d}t' \int_0^t \text{d}t'' \xi_S(t'')\ v_S(\vec{X}(t'))
    \sqrt{v_S(\vec{X}(t''))} 
  \right\rangle\quad .
\end{eqnarray}
We  show below that
\begin{eqnarray}
I(t'',t')=\left\langle \xi_S(t'')v_S(\vec{X}(t'))\sqrt{v_S(\vec{X}(t''))}\right\rangle =0\quad,  \label{eq:rel1}
\end{eqnarray}
for stationary processes. Therefore the Fano-factor equality (\ref{eq:uncertainty}) for stochastic
entropy production follows from Eq.~\eqref{eq:s3}.   To show (\ref{eq:rel1}) we first note that
for  $t''\geq t'$ the relation  (\ref{eq:rel1})  is a direct consequence of the
rules of It\^{o} calculus. This is 
because the noise in the future is uncorrelated with the trajectory in the past.
The case of  $t''<t'$ requires a careful analysis, since we have to average over the
noise $\xi_S(t'')$ conditioned on $v_S(\vec{X}(t'))$ at a future time.

To compute this conditioned average, we apply Doob's h-transform~\cite{doob1957}
(see also~\cite{satya2015,touchette}). In short, Doob's h-transform maps a stochastic
process with noise variables conditioned on a future event to a
stochastic process with unconditioned noise variables, but with an
additional drift term.
For example, consider the Langevin equation
\begin{equation}\label{eq:langevin}
\frac{{\rm d}\vec{X}}{{\rm d}t} = \bmu \cdot \vec{F}+ \vec{\nabla}\cdot \bD + \sqrt{2}\:\bsigma\cdot\vec{\xi}\quad, 
\end{equation}
with $\langle\vec{\xi}(t) \rangle=0$ and $\langle {\xi_i(t) \xi_j(t') } \rangle=\delta_{ij}\delta(t-t')$.
We calculate averages conditioned on the future constraint  $\vec{X}(t^*)=x^*$, with $t^\ast>t$.  
In other words, when taking averages we only consider the trajectories generated by (\ref{eq:langevin}) for which $\vec{X}(t^*)=\vec{x}^*$, and disregard the other ones.  In general, averages involving the noise
variables $ \xi_i(t)$ become biased by this condition, i.e., $\left.\langle \vec{\xi}(t)
\rangle\,\right|_{\vec{X}(t^\ast)=\vec{x}^\ast} \neq0$.  Introducing the $h$-function,
$h(\vec{x},t; \vec{x}^*, t^\ast)=P(\vec{x}^*,t^*|\vec{x}(t),t)$, where $P(\vec{x},t|\vec{x}_0,t_0)$ is the solution of the corresponding Fokker-Planck equation with initial condition $\vec{x}(t_0)=\vec{x}_0$.
Doob's h-transform generates a Langevin equation for a process $\vec{Z}(t)$ which reads~\cite{doob1957,satya2015,touchette}
\begin{equation}\label{eq:doobtransform}
\frac{{\rm d}\vec{Z}}{{\rm d}t} = \bmu \cdot \vec{F}+ \vec{\nabla}\cdot \bD +2
\bD\cdot\vec{\nabla}\ln h(\vec{Z},t; \vec{x}^*, t^\ast)
 + \sqrt{2}\:\bsigma\cdot\vec{\eta}\quad,   
\end{equation} 
where $\vec{\nabla}\ln h(\vec{x}_1,t_1; \vec{x}_2, t_2) =
\vec{\nabla}_{\vec{x}_1}\ln h(\vec{x}_1,t_1; \vec{x}_2, t_2)$ and $\vec{\eta}(t)$ is a white noise with zero mean, i.e.,
$\langle\vec{\eta}(t) \rangle = 0$.  Doob showed
 that  (\ref{eq:doobtransform}) generates an ensemble of
trajectories $\left\{\vec{Z}(t)\right\}_{t\in[0,t^\ast]}$  identical to the
ensemble of trajectories $\left\{\vec{X}(t)\right\}_{t\in[0,t^\ast]}$
generated by the stochastic differential equation (\ref{eq:langevin}) and
conditioned on the event $\vec{X}(t^*)=\vec{x}^*$ in the future~\cite{doob1957,satya2015,touchette}.  Comparing
Eq.~\eqref{eq:langevin} with Eq.~\eqref{eq:doobtransform}, reveals that replacing the noise in Eq.~\eqref{eq:langevin} with the noise process defined by
\begin{equation}\label{doobnoise}
\vec{\xi}(t)=\vec{\eta}(t)+\sqrt{2}\,\bsigma^{\mathsf{T}}\cdot\vec{\nabla}\ln
h(\vec{Z}(t),t;\vec{x}^\ast, t^\ast)
\end{equation} 
allows to use standard noise averages when calculating averages conditioned on the future event at time $t^{\star}$. Note that in~(\ref{doobnoise}) we have used $\bsigma\cdot\bsigma^{\mathsf{T}}=\bD$. %Note that the noise variable $\vec{\eta}(t)$ is independent of the future constraint $\vec{Z}(t^\ast) = \vec{x}^*$, since $\vec{Z}(t^\ast)=\vec{x}(t^\ast)$ takes a fixed value and is thus not a random variable.  In other words, we have $\langle \vec{\eta}(t)\vec{Z}(t^\ast) \rangle $ and $\langle \vec{\eta}(t)f(\vec{Z}(t))g(\vec{Z}(t^\ast)) \rangle = 0$ for arbitrary functions $f$ and~$g$.

Using the Doob $h$-transform, we compute now the average of
Eq.~\eqref{eq:rel1} for $t'>t''$
\begin{eqnarray}
I(t'',t')&=&\left\langle \xi_S(t'')\sqrt{v_S(\vec{X}(t''))}v_S(\vec{X}(t'))\right\rangle\nonumber\\
&=&\left\langle \frac{\vec{\xi}(t'')\cdot
    \bsigma^{-1}\cdot\vec{J}(t'')}{\sqrt{\vec{J}(t'')\cdot\bD^{-1}\cdot\vec{J}(t'')}}
\sqrt{v_S(\vec{X}(t''))}v_S(\vec{X}(t'))\right\rangle\nonumber\\
&=&\sqrt{2} \left\langle \frac{[\vec{\nabla}\ln h(\vec{Z}(t''), t'';\vec{X}(t'), t') ]
    \cdot\vec{J}(t'')}{\sqrt{\vec{J}(t'')\cdot\bD^{-1}\cdot\vec{J}(t'')}}\sqrt{v_S(\vec{Z}(t''))}v_S(\vec{X}(t'))\right\rangle\quad,
\end{eqnarray}
where $h(\vec{Z}(t''), t'';\vec{X}(t'),
t')=P(\vec{Z}(t'),t'|\vec{Z}(t''),t'')$ is the $h$-function for the future condition $Z(t') = X(t')$.  Note that we have used
$\langle f[\vec{Z}(t'),\vec{Z}(t'')] \vec{\eta}(t'')\rangle=0$, which follows from the fact that $\vec{Z}(t')$ is not fluctuating and $\vec{\eta}$ is a white noise which is uncorrelated with $\vec{Z}$ at the same time. We also used  $\bsigma\cdot\bsigma^{-1} = \mathbf{1}$. We proceed by writing the average
$\langle\dots\rangle$ explicitly in terms of the distribution of the system
states $\vec{x}'$ and $\vec{x}''$ at times $t'$ and $t''$:
\begin{eqnarray}
I(t'',t')&=& \sqrt{2\kb}\int ~\text{d}\vec{x}' ~\int~ \text{d}\vec{x}'' P(\vec{x}'', t'') P(\vec{x}',t'|\vec{x}'',t'') 
\nonumber\\ 
&& \times
\frac{\left[\vec{\nabla}_{\vec{x}''}\ln h(\vec{x}'',t'';\vec{x}', t')\right]\cdot\vec{J}(t'')}{\sqrt{\vec{J}(t'')\cdot\bD^{-1}\cdot\vec{J}(t'')}}\frac{\sqrt{\vec{J}(t'')\cdot\bD^{-1}\cdot\vec{J}(t'')}}{P(\vec{x}'', t'')}v_S(\vec{x}',t')\nonumber\\
&=&\sqrt{2\kb}\int ~\text{d}\vec{x}' ~\int~ \text{d}\vec{x}'' P(\vec{x}',t'|\vec{x}'',t'') ~\left[\vec{\nabla}_{\vec{x}''}\ln
    h(\vec{x}'',t''; \vec{x}',t')\right]\cdot\vec{J}(t'')~v_S(\vec{x}',t')\quad.\nonumber\\
\end{eqnarray}
Using $ h(\vec{x}'',t'';\vec{x}', t')=P(\vec{x}',t'|\vec{x}'',t'') $, we integrate by parts:
\begin{eqnarray}\label{doobfinal}
I(t'',t')&=& \sqrt{2\kb}\int ~\text{d}\vec{x}' ~\int~ \text{d}\vec{x}'' ~\left[\vec{\nabla}_{\vec{x}''}
P(\vec{x}',t'|\vec{x}'', t'')\right]\cdot\vec{J}(t'')~v_S(\vec{x}',t')\nonumber\\
&=&-\sqrt{2\kb}\int ~\text{d}\vec{x}' ~\int~ \text{d}\vec{x}'' ~
P(\vec{x}',t'|\vec{x}'', t'')\left(\vec{\nabla}_{\vec{x}''}\cdot\vec{J}(t'')\right)~v_S(\vec{x}',t') \nonumber\\
&=&\sqrt{2\kb}\int ~\text{d}\vec{x}' ~\int~ \text{d}\vec{x}'' ~
P(\vec{x}',t'|\vec{x}'', t'')\,\partial_{t''}
P(\vec{x}'',t'')~v_S(\vec{x}',t')\nonumber\\
&=&0\quad .
\end{eqnarray}
In the second step, we have used that no boundary term arise  
for periodic or no flux boundary conditions. Indeed  either the flux or the difference in probability must vanish for this boundary conditions. In steady state $I(t',t'')=0$ because
 $\partial_{t''} P(\vec{x}'',t'') = 0$.

\section{S3.$\quad$Fano Factor of the entropic time}

The  Fano factor of  the entropic time $\tau$ at finite times can be expressed as
\begin{equation}
\frac{\sigma^2_{\tau}}{\langle \tau\rangle}=\frac{\int_0^t \text{d}t'~
  \int_{-t'}^{t-t'} \text{d}t''~\langle v_S(\vec{X}(t'))v_S(\vec{X}(t'+t''))\rangle-\langle v_S(\vec{X}(t'))\rangle^2
}{\kb t\langle v_S\rangle}\quad,
\end{equation}
that in the limit $t\rightarrow \infty$ reduces to the Green-Kubo-like expression
\begin{equation}\label{greenkuboSI}
\frac{\sigma^2_{\tau}}{\langle \tau\rangle}=\frac{2}{\kb\langle v_S\rangle}\int_0^\infty \,\text{d}t''\, 
\langle v_S(\vec{X}(t'+t'')) v_S(\vec{X}(t'))\rangle - \langle v_S(\vec{X}(t')) \rangle^2\quad.
\end{equation}
In the following subsections, we present the calculations of the Fano
factor for the steady-state models illustrated in Fig.~\ref{fig:models} of the Main Text.

\subsection{A.$\quad$ Drift-diffusion in a triangular potential}\label{sec:triangle}

The system is defined on a one dimensional segment $[0,1]$ with
periodic boundary conditions. The potential $U(x)$ is triangular, so
that the total force is constant in the intervals $[0,x^*]$ and
$[x^*,1]$. Let us call these two regions $A$ and $B$, respectively,
and the corresponding total forces $F_A=f-\text{d} U_A(x)/\text{d}x$
and $F_B=f-\text{d} U_B(x)/\text{d}x$. In order to calculate the
long-time Fano factor Eq.~\eqref{greenkuboSI} we need to compute the
quantity
\begin{eqnarray}\label{gk_triang}
\sigma^2_\tau&=&\frac{2}{\kb^2}\int_0^\infty \,\text{d}t''\, 
\langle v_S(X(t'+t'')) v_S(X(t'))\rangle - \langle v_S(X(t')) \rangle^2
\nonumber\\
&=&\frac{2 J^4}{D^2}\int_0^\infty \text{d}t \int_0^1 \text{d}x \int_0^1 \text{d}y \frac{1}{P_{\rm st}^2(x)}
\frac{1}{P_{\rm st}^2(y)}
[P(x,t|y,0)P_{\rm st}(y)-P_{\rm st}(x)P_{\rm st}(y)]\nonumber\\
&=& \frac{2 J^4}{D^2}\int_0^\infty \text{d}t \int_0^1 \text{d}x \int_0^1 \text{d}y \frac{1}{P_{\rm st}^2(x)}
\frac{1}{P_{\rm st}(y)}
[P(x,t|y,0)-P_{\rm st}(x)]\quad,
\end{eqnarray}
where $P_{\rm st}(x)$ is the steady-state probability density. The Fokker-Planck equations read
\begin{eqnarray}\label{fp}
\partial_t P(x,t)&=&-\mu F_A\partial_x P +D\partial_x^2 P
\qquad x\in [0,x^*]\nonumber\\
\partial_t P(x,t)&=&-\mu F_B\partial_x P +D\partial_x^2 P  
\qquad x \in [x^*,1]\quad.
\end{eqnarray}
Let us first compute $P_{\rm st}(x)$. Solving Eq. (\ref{fp}) at steady state yields
\begin{eqnarray}\label{pst}
P_{\rm st}(x)&=&\alpha_1+\alpha_2 e^{F_A x/(\kb T)} \qquad x \in [0,x^*] \nonumber\\
P_{\rm st}(x)&=&\alpha_3+\alpha_4 e^{F_B x/(\kb T)} \qquad x \in [x^*,1]\quad.\label{eq:psttriangle}
\end{eqnarray}
The four integration constants are determined by imposing 1) normalization of
$P_{\rm st}(x)$, 2) conservation of current $J_A=J_B$, 3) continuity of
probability in $x=x^*$, and 4) periodic boundary condition
$P_{\rm st}(0)=P_{\rm st}(1)$. Notice that all these conditions are linear in the
integration constants. The explicit solution is
\small
\begin{eqnarray}
\alpha_1 &=& \frac{F_A F_B^2\left[e^{F_B x^*/(\kb T)}-e^{(F_B+F_A x^*)/(\kb T)}\right]}{\mathcal{N}} \nonumber\\
\alpha_2 &=& \frac{F_A F_B(F_B-F_A)\left[e^{F_B/(\kb T)}-e^{F_B x^*/(\kb T)}\right]}{\mathcal{N}} \nonumber\\
\alpha_3 &=& \frac{F_A^2 F_B\left[e^{F_B x^*/(\kb T)}-e^{(F_B+F_A x^*)/(\kb T)}\right]}{\mathcal{N}} \nonumber\\
\alpha_4 &=&  \frac{F_A F_B(F_A-F_B)\left(e^{F_A x^*/(\kb T)}-1\right)}{\mathcal{N}} \quad,
\label{eq:alphasttriangle}
\end{eqnarray}
with the normalization constant $\mathcal{N}=\kb T (e^{F_A
    x^*/(\kb T)}-1)(e^{F_B/(\kb T)}-e^{F_B x^*/(\kb T)})(F_A-F_B)^2+
F_A F_B\left[e^{F_B x^*/(\kb T)}-e^{(F_B+F_A x^*)/(\kb T)}\right](F_A-F_Ax^*+F_Bx^*)$.

Note that 
\begin{equation}
v_S(X(t)) = \frac{\kb J^2}{DP_{\rm st}^2(X(t))}\quad,
\end{equation}
with the current given by $J=\mu F_A \alpha_1=\mu F_B \alpha_3$, and $P_{\rm st}(x)$ given by Eqs.~(\ref{eq:psttriangle}) and~(\ref{eq:alphasttriangle}). Its average equals
\begin{equation}
\langle v_S\rangle=\frac{\kb J^2}{D}\left\langle\frac{1}{P_{\rm st}^2} \right\rangle=\frac{\kb J^2\gamma}{D}\quad,
\end{equation}
where we have defined the quantity
\begin{eqnarray}
\gamma &= &\int_0^1 \text{d}x \frac{1}{P_{\rm st}(x)}  \nonumber\\
&=& \frac{x^*}{\alpha_1}+
\kb T \frac{\ln(\alpha_1+\alpha_2)
-\ln\left(\alpha_1+\alpha_2 e^{F_A x^*/D}\right)}{\alpha_1 F_A}\nonumber\\
&+& \frac{1-x^*}{\alpha_3}+
\kb T\frac{\ln(\alpha_3+\alpha_4 e^{F_B x^*/D})-\ln\left(\alpha_3+\alpha_4 e^{F_B /D}\right)}{\alpha_3 F_B}\quad.
\end{eqnarray}

Let us now define
\begin{equation}
f(x,t)=\int_0^1 \text{d}y\ \frac{P(x,t|y,0)-P_{\rm st}(x,t)}{P_{\rm st}(y)}\quad.
\end{equation}
The function $f(x,t)$ is a solution of the Fokker-Planck Eq. (\ref{fp}) with
initial condition $f(x,0)=1/P_{\rm st}(x)-\gamma P_{\rm st}(x)$. Notice that $f(x,t)$
is not a probability distribution as $\int \text{d}x f(x,t)=0$ $\forall t$. We also
introduce an additional function $\phi(x)=\int_0^\infty \text{d}t f(x,t)$. Integrating the Fokker-Planck equation in time, we find that
$\phi(x)$ obeys 
\begin{eqnarray}
-\mu F_A\partial_x \phi +D\partial_x^2 \phi &=& -f(x,0)=-1/P_{\rm st}(x)+\gamma P_{\rm st}(x)
\qquad x\in [0,x^*]\nonumber\\
-\mu F_B\partial_x \phi +D\partial_x^2 \phi&=&-f(x,0)=-1/P_{\rm st}(x)+\gamma P_{\rm st}(x)
\qquad x \in [x^*,1]\quad.
\end{eqnarray}
The solution to this equation is 
\begin{eqnarray}
\phi(x)&=&\beta_1+\frac{1}{\alpha_1^2 \mu F_A^2}\left\{   
e^{F_A x/(\kb T)}\left[\alpha_1^2\beta_2 \mu\kb T F_A+
           \alpha_1^2\alpha_2 \gamma (F_A
  x-\kb T)
-\alpha_2 \kb T\ln\left(\alpha_2+\alpha_1 e^{-F_Ax/(\kb T)}\right)\right]\right.\nonumber\\
&-&\left.\alpha_1(\gamma \alpha_1^2-1)F_A
    x-\kb T\alpha_1\ln\left(\alpha_1+\alpha_2e^{F_A x/(\kb T)} \right)
\right\} \qquad x\in [0,x^*]\nonumber\\
\phi(x)&=&\beta_3+\frac{1}{\alpha_3^2 \mu F_B^2}\left\{   
e^{F_B x/\kb T}\left[\alpha_3^2\beta_4 \mu \kb T F_B+ \alpha_3^2\alpha_4 \gamma (F_B x-\kb
  T)
-\alpha_4 \kb T   \ln\left(\alpha_4+\alpha_3 e^{-F_Bx/(\kb T)}\right)\right]       \right.\nonumber\\
&-&\left.\alpha_3(\gamma \alpha_3^2-1)F_B
    x-\kb T\alpha_3\ln\left(\alpha_3+\alpha_4e^{F_B x/(\kb T)} \right)
\right\} \qquad x\in [x^*,1]\quad.
\end{eqnarray}
The four integration constants $\beta_i$, $i=1\dots 4$ can be determined using
similar conditions we imposed for the stationary distribution~(\ref{pst}). %Using the fact that the function $\phi$ is normalized to zero and the integrated current is continuous. All these conditions are again linear.

Defining the quantity $\psi=\int_0^1 \text{d}x\ \phi(x)/P^2_{\rm st}(x)$,
Eq.~\eqref{gk_triang} can be rewritten as
\begin{equation}\label{gk_triang2}
\frac{\sigma^2_\tau}{\langle\tau\rangle}=\frac{2J^2\psi}{D\kb\gamma}\quad .
\end{equation}
The integral in the definition of $\psi$ can be computed analytically,
but yields a lengthy expression
involving special functions that is hard to evaluate numerically. For
this reason, the theoretical line in Fig.~\ref{fig:fano} of the Main Text was
obtained from Eq.~\eqref{gk_triang2} by numerically integrating the
expression for $\psi$.

\subsection{B.$\quad$2D transport in a force field}\label{sec:2d}

We consider the following two-dimensional dynamics coordinate
\begin{eqnarray}
\frac{\text{d}{X}}{\text{d}t}&=&\mu F(Y)+\sqrt{2D} \xi_X \nonumber\\
\frac{\text{d}{Y}}{\text{d}t}&=&\sqrt{2D} \xi_Y  \quad,
\end{eqnarray}
with the non-conservative force $F(y)=f \cos(2\pi y)$. This case is
considerably simpler than that of the previous section as the stationary
distributon is homogeneous, so that 
\begin{equation}
v_S (X(t),Y(t)) =v_S (Y(t)) = \frac{\kb \mu^2 F^2(Y(t))}{D} \quad,
\end{equation}
and its average its given by
\begin{equation}
\langle v_S\rangle = \kb\int_0^1\frac{\mu^2 F^2(y)}{D} \text{d}y =\frac{\mu
  f^2}{2T}\quad.
\end{equation}
Moreover, since the force depends only on the $y$ position, the correlation entering the Green-Kubo formula can be calculated  and gives
\begin{eqnarray}
  \langle\, v_S(y(0)) v_S(y(t))\,\rangle & = &\frac{\kb^2\mu^4}{D^2} 
  \int_0^1 dy \int_{-\infty}^\infty
  d \Delta y\ F^2(y)\ F^2(y+\Delta y) \frac{e^{-\frac{\Delta
        y^2}{4Dt}}}{\sqrt{4\pi D t}}\nonumber\\
  &=& \frac{\mu^2f^4}{4T^2} \left[1+\frac{1}{2}e^{-16\pi^2 D t}\right]\quad.
\end{eqnarray}
Substituting this expression in Eq.~\eqref{greenkuboSI}
we obtain
\begin{equation}
\frac{\sigma^2_\tau}{\langle \tau\rangle} = \frac{ f^2 }{32\pi^2\kb^2 T^2}\quad.
\end{equation}

\subsection{C.$\quad$Chiral active Brownian motion}\label{sec:3d}

This model is defined by the set of equations
\begin{eqnarray}
\text{d}X/\text{d}t&=&\mu f \cos(\phi)+\sqrt{2D}\xi_x\nonumber\\
\text{d}Y/\text{d}t&=&\mu f \sin(\phi)+\sqrt{2D}\xi_y \nonumber\\
\text{d}\phi/\text{d}t&=&\mu_\phi \omega+\sqrt{2D_\omega}\xi_\omega \quad.
\end{eqnarray}
In this case, the entropy drift is constant and given by
\begin{equation}
v_S=\frac{\kb\mu^2 f^2}{D}+\frac{\kb\mu^2_\phi
  \omega^2}{D_\phi}\quad.
  \end{equation}
Therefore $\tau$ grows in a deterministic
way, $\sigma^2_{\tau}=0$, the distribution of $S_{\rm tot}$ is Gaussian and the Fano factor of $S_{\rm tot}$ is equal to 2.

\section{S4.$\quad$Fano factor of entropy production out of steady state}\label{sec:fanoout}
In this section, we derive the Fano factor equality out of steady state, given
by Eq.~\eqref{eq:uncertainty2} in the main text.  We find the average of the squared entropy
production from the expression for the stochastic entropy production
Eq.~(\ref{eq:entropySI}), namely,
\begin{eqnarray}\label{eq:S2_out}
  \frac{\langle S^2_{\rm tot}(t)\rangle}{\kb^2} 
&=&\left\langle \left[\int_0^t\ \text{d}t' \left\{
  \frac{v_S(\vec{X}(t'),t')}{\kb}-2\partial_{t'} \ln P(\vec{X}(t'),t') +\xi_S(t')\ \sqrt{\frac{2v_S(\vec{X}(t'),t')}{\kb}}\right\} \right]^2\right\rangle 
\nonumber\\
  &=&\langle \tau^2 \rangle+2\langle \tau\rangle +4 \left\langle\left[\int_0^t
    \text{d}t'\ \partial_{t'} \ln P(\vec{X}(t'),t')\right]^2   \right\rangle        
-4 \left\langle\tau(t)\int_0^t
    \text{d}t'\ \partial_{t'} \ln P(\vec{X}(t'),t')   \right\rangle 
  \nonumber\\ 
  &+&2\sqrt{2}\left\langle\int_0^t \text{d}t' \int_0^t \text{d}t'' \xi_S(t'')\ 
    \sqrt{\frac{v_S(\vec{X}(t''),t'')}{\kb}} \left[\frac{v_S(\vec{X}(t'),t')}{\kb}-2\partial_{t'}\ln P(\vec{X}(t'),t')\right]
\right\rangle \quad. \nonumber\\
\end{eqnarray}
We  first evaluate the term in the last line using Doob's h-transform, as in  
Eqs.~\eqref{eq:rel1}-\eqref{doobfinal}.   Following this procedure, we obtain
\begin{eqnarray}
  \lefteqn{2\sqrt{2} \int_0^t \text{d}t' \int_0^t \text{d}t''  \left\langle \xi_S(t'')\ 
      \sqrt{v_S(\vec{X}(t''),t'')} \left[\frac{v_S(\vec{X}(t'),t')}{\kb^{3/2}}-
\frac{2}{\sqrt{\kb}}\partial_{t'}\ln P(\vec{X}(t'),t')\right]\right\rangle}&&
  \nonumber\\
  &=&4  \int_0^t \text{d}t' \int_{0}^{t'} \text{d}t''  \left\langle~\partial_{t''}
    \ln P(\vec{X}(t''),t'')~\left[\frac{v_S(\vec{X}(t'),t')}{\kb}-2\partial_{t'}\ln P(\vec{X}(t'),t')\right]   \right\rangle\quad .
\end{eqnarray}
Substituting this expression into Eq.~\eqref{eq:S2_out} yields
\begin{eqnarray}\label{eq:S2_out2}
  \frac{\langle S^2_{\rm tot}(t)\rangle}{\kb^2}& =&\langle \tau^2(t)
  \rangle+2\langle \tau(t)\rangle- \frac{4}{\kb} \int_0^t \text{d}t' \int_{0}^{t'} \text{d}t'' ~ \left\langle  \partial_{t''}
  \ln P(\vec{X}(t''),t'')~v_S(\vec{X}(t'),t')\right \rangle \quad,
\end{eqnarray}
which leads to Eqs.~\eqref{eq:uncertainty2},~\eqref{eq:omega}, and~\eqref{eq:F} in the Main Text.

\section{S5.$\quad$Fano Factor of Entropy production for a quench of an harmonic
trap}

In this section, we briefly present details of the numerical
simulations shown in Fig.~\ref{fig:quench} of the Main Text. We consider a system described by the one dimensional Fokker-Planck equation
\begin{equation}\label{eq:ou}
\partial_t P(x,t)=\partial_x[\mu \kappa_f xP(x,t)+D\partial_x P(x,t)]\quad.
\end{equation}
The system is initially prepared at thermal equilibrium with a stiffness
$\kappa_i$ and then instantaneously quenched to the stiffness $\kappa_f$.
The Fano factor of total entropy production at
long times, shown in Fig.~\ref{fig:quench} in the Main Text,  is computed from simulations by the usual definition of
total entropy production $S_{\rm tot}=(W-\Delta F)/T$ so that
\begin{equation}\label{fanoquench}
\frac{\sigma^2_{S_{\rm tot}}}{\kb\langle S_{\rm tot}\rangle}=\frac{\sigma^2_W}{\kb T(\langle
  W\rangle-\Delta F)}\quad.
\end{equation}
In this case, the  work is simply given by the instantaneous change in energy due to the
quench
\begin{equation}
W=\left(\frac{\partial U}{\partial k}\right)\Delta k =
\frac{1}{2}(\kappa_f-\kappa_i)x^2(t=0)\quad.
\end{equation}
Averaging over the initial condition, the mean work is
\begin{equation}
\langle W\rangle =\int_{-\infty}^{\infty} \text{d}x~P(x,0) \frac{1}{2}(\kappa_f-\kappa_i)x^2 =\frac{\kb T}{2}\frac{(\kappa_f-\kappa_i)}{\kappa_i} \quad,
\end{equation}
 where we used the fact that $P(x,0)$ is Gaussian with mean $0$ and variance $\kb T/\kappa_i$. Similarly we have
\begin{equation}
\langle W^2 \rangle =\int_{-\infty}^{\infty} \text{d}x\,P(x,0)\left(\frac{1}{2}(\kappa_f-\kappa_i)x^2\right)^2 =\frac{3(\kb  T)^2}{4}\frac{(\kappa_f-\kappa_i)^2}{\kappa_i^2} \quad,
\end{equation}
so that 
\begin{equation}
\sigma^2_W=\langle W^2 \rangle-\langle W\rangle^2=\frac{(\kb
  T)^2}{2}\frac{(\kappa_f-\kappa_i)^2}{\kappa_i^2}  \quad.
\end{equation}
 Finally, for this protocol we simply have 
\begin{equation}
\Delta F=\frac{\kb T}{2}\log(\kappa_f/\kappa_i)\quad .
\end{equation}
Plugging these expressions into Eq.~\eqref{fanoquench} we obtain an
exact expression for the Fano factor:
\begin{equation}\label{fanoquench}
\frac{\sigma^2_{S_{\rm tot}}}{\kb\langle S_{\rm tot}\rangle}=\frac{(\kappa_f/\kappa_i-1)^2}{(\kappa_f/\kappa_i-1)-\log(\kappa_f/\kappa_i)}\quad.
\end{equation}

To compare Eq.~(\ref{fanoquench}) with Eq.~\eqref{eq:uncertainty2} in the
Main Text we recall that Eq.~\eqref{eq:ou} describes an Ornstein-Uhlenbeck
process, for which the propagator reads
\begin{equation}\label{propagator}
P(x,t|x',t')=\sqrt{\frac{\kappa_f}{2\pi\kb T(1-e^{-2\mu
      \kappa_f(t-t')})}}\exp\left[-\frac{\kappa_f(x-e^{-\mu
      \kappa_f(t-t')}x')^2}{2\kb T(1-e^{-2\mu
      \kappa_f(t-t')})} \right]\quad.
\end{equation}
Notice that the propagator is Gaussian.
The system is initially in equilibrium with a
different stiffness $\kappa_i$, so that its initial distribution is Gaussian with
mean zero and variance $\sigma^2_i=\kb T/\kappa_i$. Integrating Eq.~\eqref{propagator} over such
initial condition, we find that the distribution of the system remains Gaussian at all
times:
\begin{equation}
P(x,t)=\frac{1}{\sqrt{2\pi\sigma^2(t)}}\exp[-x^2/2\sigma^2(t)]\quad,
\end{equation}
where
\begin{eqnarray}
\sigma^2(t)&=&\sigma^2_f+(\sigma_i^2-\sigma_f^2)e^{-2\mu \kappa_f t}\quad,
\end{eqnarray}
and $\sigma^2_f=\kb T/\kappa_f$. 
We can now write the probability current as
\begin{eqnarray}
J(x,t) &=& -\mu \kappa_f x  P(x,t) - D \partial_x P(x,t) \quad.
\end{eqnarray}
The corresponding entropic drift is
\begin{eqnarray}\label{vsquench}
\frac{v_S(X(t),t)}{\kb}&=&\frac{J^2(X(t),t)}{DP^2(X(t),t)}\nonumber\\
&=& \left( \frac{\kappa_f}{\kappa_i}-1\right)^2 e^{-4\mu \kappa_f t}\frac{DX^2(t)}{\sigma^4(t)}\quad.
\end{eqnarray}
Similarly, one can show that
\begin{eqnarray}\label{lnpquench}
\partial_t \ln P(x,t)=\mu \kappa_f [x^2-\sigma^2(t)]
  \frac{(\sigma^2_f-\sigma^2_i) e^{-2\mu \kappa_f t}}{\sigma^4(t)}\quad.
\end{eqnarray}
Using Eqs.~\eqref{vsquench} and \eqref{lnpquench}, one can
estimate the quantities $\sigma^2_\tau$ and $\Omega$ by calculating suitable
averages  in numerical simulations of the process.  The curves in Figs.~\ref{fig:quench}a and~\ref{fig:quench}b in the Main Text have
been computed by this method.

 %\newpage \newpage
%\includepdf[pages={1,{},{},2,{},3,{},4,{},5,{},6,{},7,{},8,{},9,{},10,{},11}]{supplements_arxiv_v2.pdf}

\end{document}